\documentclass[twocolumn]{aastex631}

\shorttitle{HD 163296: crescent and resonant chain}
\shortauthors{Garrido-Deutelmoser, et al.}

\begin{document}

\title{A gap-sharing planet pair shaping the crescent in HD 163296: a disk sculpted by a resonant chain}

\author[0000-0002-7056-3226]{Juan Garrido-Deutelmoser}
\affiliation{Instituto de Astrofísica, Pontificia Universidad Cat\'olica de Chile, Av. Vicuña Mackenna 4860, 782-0436 Macul, Santiago, Chile}
\affiliation{N\'ucleo Milenio de Formaci\'on Planetaria (NPF), Chile}

\author[0000-0003-0412-9314]{Cristobal Petrovich}
\affiliation{Instituto de Astrofísica, Pontificia Universidad Cat\'olica de Chile, Av. Vicuña Mackenna 4860, 782-0436 Macul, Santiago, Chile}
\affiliation{Millennium Institute for Astrophysics, Chile}

\author[0000-0002-9196-5734]{Carolina Charalambous}
\affiliation{naXys, Department of Mathematics, University of Namur, Rue de Bruxelles 61, 5000 Namur, Belgium}

\author[0000-0003-4784-3040]{Viviana V. Guzm\'an}
\affiliation{Instituto de Astrofísica, Pontificia Universidad Cat\'olica de Chile, Av. Vicuña Mackenna 4860, 782-0436 Macul, Santiago, Chile}
\affiliation{N\'ucleo Milenio de Formaci\'on Planetaria (NPF), Chile}

\author[0000-0002-0661-7517]{Ke Zhang}
\affiliation{Department of Astronomy, University of Wisconsin-Madison, 475 N. Charter Street, Madison, WI 53706, USA}



\begin{abstract}

 ALMA observations of the disk around HD 163296 have resolved a crescent-shape substructure at around 55 au, inside and off-center from a gap in the dust that extends from  38 au to 62 au. In this work we propose that both the crescent and the dust rings are caused by a compact  pair (period ratio $\simeq 4:3$) of sub-Saturn-mass planets inside the gap, with the crescent corresponding to dust trapped at the $L_5$ Lagrange point of the outer planet. This interpretation also reproduces well the gap in the gas recently measured from the CO observations, which is shallower than what is expected in a model where the gap is carved by a single planet.  Building on previous works arguing for outer planets at $\approx 86$ and $\approx 137$ au, we provide with a global model of the disk that best reproduces the data and show that all four planets may fall into a long resonant chain, with the outer three planets in a 1:2:4 Laplace resonance. We show that this configuration is not only an expected outcome from disk-planet interaction in this system, but it can also help constraining the radial and angular position of the planet candidates using three-body resonances.

\end{abstract}


\keywords{protoplanetary disks --- planet–disk interactions ---  hydrodynamics --- planets and satellites: dynamical evolution and stability --- radiative transfer}


\section{Introduction} \label{sec:intro}

Substructures are ubiquitous in protoplanetary disks, particularly in the dust density distribution exhibited by high angular resolution observations \citep{Andrews2020,Bae2022_PPVI}. The Atacama Large Millimeter Array (ALMA) has revealed a variety of substructures, whereas a large population of rings and gaps are shown in continuum observations, to a lesser extent, in molecular line emissions (e.g., \citealt{vanderMarel2019}).

The advances in spatial resolution have been able to resolve non-axisymmetric substructures within gaps, including systems such as PDS 70 \citep{Benisty2021}, HD 163296 \citep{Isella2018}, HD 100546 \citep{Perez2020}, HD 97048 \citep{Pinte2019}, and LkCa 15 \citep{Feng2022}. These substructures may be due to embedded planets induced by gravitational interactions (e.g., \citealt{Bae2022_PPVI}), which vary depending on the emission shape. Point-like emissions are generally associated with an accreting planet surrounded by a circumplanetary disk (CPD) \citep{Perez2015,Szulagyi2018}, while crescent shapes may be  related to the stable Lagrange points $L_4$ and $L_5$ of a star-planet system \citep{Rodenkirch2021,Feng2022} or vortices. These hypotheses frequently assume that a single substructure is caused by a single planet, often in a Jovian mass regime. However, we have recently shown in \citet{Garrido-Deutelmoser}, that a pair of lower-mass and gap-sharing planets can sculpt compact and/or elongated vortices within the gap that last for several thousands orbits.

The disk surrounding HD 163296 contains ringed structures in the mm-continuum \citep{Isella2018} and several molecular tracers \citep{Law2021, Zhang2021}. In particular, inside the dust density gap that extends from 38 to 62 au, a crescent-shaped substructure resides at around 55 au \citep{Teague2021}. Recently, it was suggested that the emission comes from the dust trapped around the stable point $L_5$ of a Jupiter mass planet orbiting at 48 au \citep{Rodenkirch2021}. Even though this method seems to reproduce the broad features of the dust continuum distribution, two aspects remain unclear:

\begin{enumerate}
\item the crescent feature resides at 55 au, off-center from the dust gap, while the $L_5$ point is co-orbital to the planet at 48 au. Varying the planet's eccentricity to account for this shift is unlikely to help as the stability of the crescent is damaged, leading to its prompt disruption. 

\item the Jupiter-mass planet needed to retain observable amounts of dust at $L_5$ and open a wide enough gap in the dust, is expected to  open a deep gap\footnote{An increase in the local viscosity to produce a much shallower gap comes at the expense of reducing the lifetime of the $L_5$ crescent, or even prevent its formation in the first place \citep{Rodenkirch2021}.} \citep{Duffell2020}. This prediction disagrees with the recent results provided by \citet{Zhang2021}, who found that the dust density gap has a corresponding CO gap $\sim$ 10 times shallower than the predictions involving a Jupiter. The local gas depletion depends on the planetary mass ($\propto m_{\rm p}^{-2}$) \citep{Kanagawa2015}, whereby opting for a lower mass planet to carve a shallower gap, may not produce a sufficient gravitational interaction to enforce the dust trapping at $L_5$.
\end{enumerate}

In this work, we propose a that a compact pair of sub-Saturn-mass planets can solve these issues, simultaneously accounting for the dust emission (the shifted crescent and dust rings) and shallow gap in the CO. This scenario is largely motivated by our recent work in \citet{Garrido-Deutelmoser} where we showed that a compact pair of gap-sharing planets generally lead to  nonaxisymmetric substructures like that observed in HD 163296. 

\section{Setup}\label{sec:2_setup}

The hydrodynamics simulations and radiative transfer calculations in this work largely follow the scheme in \citet{Rodenkirch2021} and are only briefly summarized here. We carried out 2D hydrodynamic simulations using the \textsc{FARGO3D} multifluid code \citep{Benitez2019,Masset2000,Weber2019} to produce gas and dust density distribution for a fiducial disk model. The resulting density maps are read into the \textsc{RADMC3D} code \citep{Dullemond2012} to calculate the radiative transfer image at $\lambda \sim$ 1.25 mm. We use this image and the HD 163296 template (that contain all the technical properties of the observation) in the \textsc{SIMIO}\footnote{\href{https://www.nicolaskurtovic.com/simio}{https://www.nicolaskurtovic.com/simio}} package to achieve the synthetic ALMA observation comparable to the dust continuum observation provided by \citet{Isella2018}.

\subsection{Hydrodynamic Simulations}
\label{sec:hydro_sims}
The initial surface density profiles for the gas (sub-index g) and dust (sub-index d) are given by
\begin{equation}
    \Sigma_{\rm g/d} = \Sigma_{{\rm g/d,0}} \left(\frac{r}{r_0}\right)^{-0.8} {\rm exp}\left[-\left(\frac{r}{r_{\rm c,g/d}}\right)^{\gamma_{\rm g/d}}\right],
\end{equation}
where we set $r_0 = 48$ au, the initial surface density $\Sigma_{\rm g,0} = 37.4 \;\rm gr\;cm^{-2}$, the cut-off radius $r_{\rm c,g} = 165$ au, and the exponent $\gamma_{\rm g} = 1$. Similarly, for the dust we set $\Sigma_{\rm d,0} = \Sigma_{\rm g,0}/100 = 0.374 \;\rm gr\;cm^{-2}$, $r_{\rm c,d} = 90$ au, and $\gamma_{\rm d} = 2$. We include the evolution for five independent dust species. These grains have sizes in cm of $0.02$, $0.071$, $0.13$, $0.26$, and $1.92$. We use an aspect ratio of $h(r) = h_0(r/r_0)^f$ with $h_0=0.05$ and a flaring index $f=0.25$, which implies a mid-plane temperature profile described by $T = 25 (r/r_0)^{-0.5}$ K. This setup coincides with that from \citet{Rodenkirch2021}.

The disk extends from $r_{\rm in} = 5$ au to $r_{\rm out} = 197$ au, implying an initial disk mass of $\sim 0.15\; M_{\rm \odot}$. The computational domain is composed of $n_{\rm r} = 512$ logarithmically spaced radial cells and $n_{\rm \theta} = 768$ equally spaced cells in the azimuthal $[0,2\pi]$ domain. We include a radially variable viscosity of the standard parameter $\alpha$ \citep{Shakura1973} as
\begin{equation}
    \alpha(r) = \alpha_{\rm in}-\frac{\alpha_{\rm in}-\alpha_{\rm out}}{2}
    \left[1+{\rm tanh}\left(\frac{r-\xi}{\sigma r_0}\right)\right],
\end{equation}
where the inner and outer viscosities are $\alpha_{\rm in} = 1\times10^{-4}$ and $\alpha_{\rm out} = 5\times10^{-3}$, $\xi=144$ au indicate the midpoint of the transition and $\sigma=1.25$ defines the slope. Similar to the values provided by \citet{Liu2018}.

A system of 4 planets was embedded. The location of the two outer ones is indicated in \citet{Teague2018} through kinematic detections. The third planet's position (i.e., the inner planet of the outer pair) is strongly associated with the potential velocity kink reported by \citep{Pinte2020}. The two inner planets are tightly packed and their parameters (masses and orbits) were derived numerically  guided by the results in \citet{Garrido-Deutelmoser}, where it was found that the planets should be gap-sharing with forming vortices at their Lagrange points, implying a condition in the planetary separation\footnote{This expression has been tested for planets with masses near the thermal mass of $M_{\rm th}=M_\star h^3\sim0.25M_{\rm J}\simeq 80M_\oplus$.} of:
\begin{eqnarray}
\Delta a\lesssim 4.6 H\simeq 11.5\mbox{ au},
\end{eqnarray}
where $H$ the scale high of the disk. In turn, the masses are constrained by the width of the gap. After a few dozen simulations attempting to match disk morphology in continuum observations at the $\sim 48$ au region, we choose to place the planets at $a_1=46$, $a_2=54$, $a_3=84.5$, and $a_4=137$ au with their respective masses of $M_1=85 M_{\oplus}$, $M_2=60 M_{\oplus}$, $M_3=0.4 M_{\rm Jup}$, and $M_4=1 M_{\rm Jup}$. The four bodies can gravitationally interact between them, but they do not feel the disk. We ran an extra model to compare against previous works, substituting a Jupiter at 48 au instead of the inner package of planets. Both cases were evolved for $0.48$ Myrs, equivalent to 2000 orbits of the innermost planet.

\subsection{Radiative Transfer}

We convert the 2D dust surface density into a 3D volume density assuming the vertical approximation given by
\begin{equation}
    \rho_{{\rm d}_j}(r,\phi,\theta) =  \frac{\Sigma(r,\phi)}{\sqrt{2 \pi}H_{{\rm d}_j}} {\rm exp}\left(-\frac{z^2}{2 H_{{\rm d}_j}^2}\right),
\end{equation}
where  $z = r \cos(\theta)$ and the dust settling follows the diffusion model $H_{{\rm d}_j} = \sqrt{D_z/(D_z+{\rm St}_j)}H$, with $D_z = 0.6 \alpha$ the vertical diffusion coefficient, and ${\rm St}_j$ the Stokes number of the species  $j$ \citep{Weber2022}. We assume an intrinsic volume density for the particles $\rho_s = 2 \;\rm gr\;cm^{-3}$ and a power law for the grain size distribution, such that $n(a)\propto a^{-3.5}$. We assumed a dust composition of 20$\%$ amorphous carbon, 20$\%$ water ices and 60$\%$ silicates, where the corresponding dust opacities were computed with the code provided by \citep{Bohren1983}. The polar direction is distributed in 64 equally spaced cells and extended in [$80.6^{\circ}, 99.4^{\circ}$] inclination domain. We use $n_{\rm phot} = 10^8$ photon packages to calculate the dust temperature, and $n_{\rm scatt} = 10^7$ photon packages to trace the thermal emission. We use a full anisotropic scattering with polarization treatment. The system is assumed to be at a distance of $101$ pc with a central star of mass $1.9$ $M_{\odot}$ and effective temperature $T_{\rm eff} = 9,330$ K. The inclination is taken to be $i=46^{\circ}$ and position angle PA$=133^{\circ}$.

\subsection{Synthetic Observations}

We use \textsc{SIMIO} that contains a suit of functions for \textsc{CASA 5.6.2}. We select the template designed for HD 163296 to create images with the same $uv$-coverage as observation from \citet{Isella2018}. We set the rescale flux option in 0.4 to get similar intensities. In addition, we add simple thermal noise\footnote{\url{https://simio-continuum.readthedocs.io/en/main/tutorials/tutorial_3.html}} of level $12 {\rm mJy}$ to finally get RMS noise of $0.022 \;{\rm mJy \;beam^{-1}}$.

\section{Local gap depletion}

\begin{figure*}[t]
    \centering
    \includegraphics[width=0.95\linewidth]{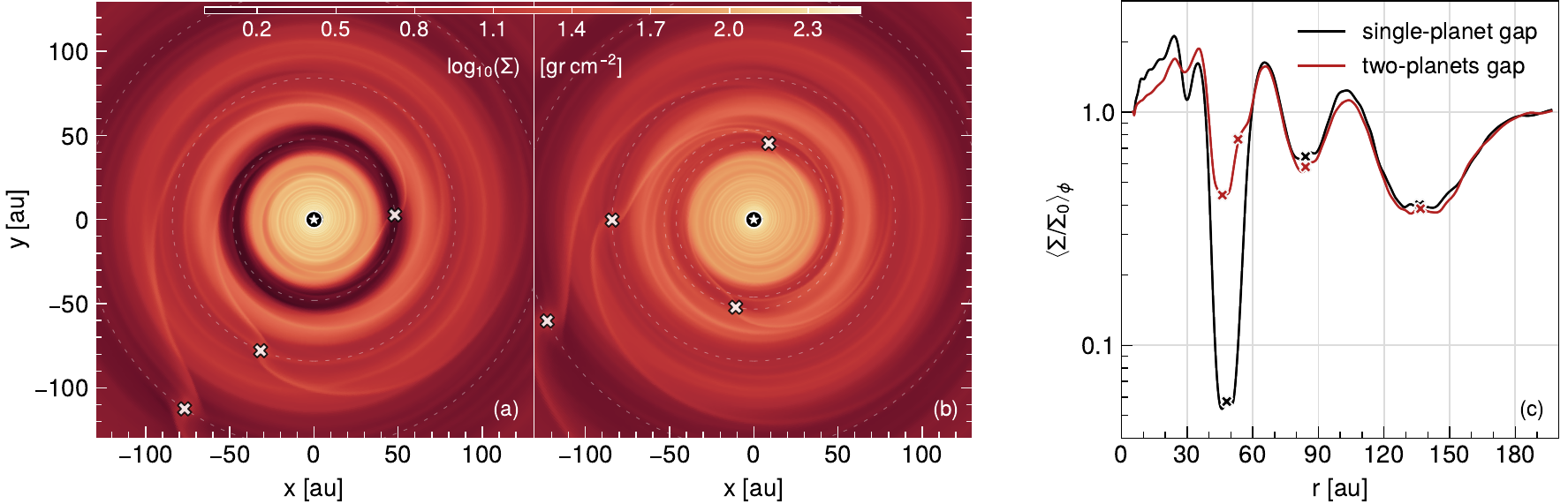}
    \caption{Panels show a time evolution after 2000 orbits at 48 au ($\sim 4.8\times 10^{5}$ yrs). (a) Surface density $\Sigma$ maps in log-scale for a single Jupiter planet at 48 au. (b) The same as (a), but for a two-planet system with $85 M_{\oplus}$ and $60 M_{\oplus}$ instead the Jupiter. The crosses denote the position of the planets and the white lines indicate their orbits. The disk rotates in a clockwise direction. (c) Azimuthally averaged $\Sigma/\Sigma_0$ profiles for both models. }
    \label{fig:Depth_Comparison}
\end{figure*}

Figure \ref{fig:Depth_Comparison} shows the surface density after $\sim 0.5$ Myr (2,000 orbits) for the single-Jovian case (panel a) and the two sub-Saturns with masses $85 M_{\oplus}$ and $60 M_{\oplus}$ (panel b). The corresponding azimuthally-averaged profiles in  panel (c) show that $\sim 95 \%$ of gas is depleted for the single-Jovian, while only $\sim 55 \%$ is depleted for the two-planet case. Despite of their lower masses the planets pair creates the same gap width as the Jovian. This shallower gaps for fixed gap width are expected in compact multi-planet systems due the planet lower masses (depth $\Sigma_{\rm gap}/\Sigma_0\propto M_p^{-2}$) and angular flux transferred by the neighbouring planets \citep{duffell_dong2015,Garrido-Deutelmoser}.

As argued by \citet{Zhang2021}, if a Jupiter-mass planet had opened the corresponding CO gap, it would be $10$ times deeper than what is actually observed. Instead, by embedding two planets we can alleviate these differences and largely reduce this discrepancy  as shown in panel (c). Therefore, the depletion values would be closer to the results from observations. In order to better quantify this, we compare the $\rm CO$ column density gaps with the surface density from both models\footnote{The $\Sigma$ profiles were processed under smooth function methodology described described in Appendix \ref{sec:appendix_fit}.}. As shown in Figure \ref{fig:CO_Comparison}, this approach reproduces the gas gap reasonably well in the two-planet case, largely matching the depths and widths with a small offset of the peaks by $\sim 4$ au. In turn, the single-Jovian case is too deep compared to the observations as expected.

Beyond the third planet at $\sim 85 $ au, neither of the two models (single Jovian or compact pair) is able to reproduce the gas gaps. This was already noted in \citet{Zhang2021} when comparing with the models from \citet{Teague2021} and it may partly be explained by the presence of a CO snowline at 65 au. Outside the mid-plane CO snowline, CO freezes out at the disk mid-plane and therefore the CO gap properties (e.g., width and depth) may deviate from that of gas gaps due to vertical temperature and CO abundance variations. We recall that our work mostly focuses on the gap and crescent at $\sim 50$ au where these issues can be more securely avoided.

\section{The Crescent Feature}

A closely-packed planet pair can directly affect the gas around each other's coorbital regions by depositing angular momentum from wave steepening and subsequent shocks. These shocks may strengthen the vortencity around the stable $L_4$ and $L_5$ Lagrange points, thus enabling the effective gas and dust trapping for at least thousands of orbital periods \citep{Garrido-Deutelmoser}. 

The overdensity around either $L_4$ or $L_5$ for the inner and outer planet is a highly dynamic problem, with the most prominent structure chaotically alternating location. This said, we observe that for several combinations of parameters (surface density, aspect ratio, planetary masses, and so on), the outer planet often retains large amounts of material around $L_5$. The depicted behavior would leave an off-center substructure inside the gap that greatly reproduces the distinctive emission in the disk as shown in Figure \ref{fig:ALMA_Comparison}. More quantitatively, our models matches the observed azimuthal extent of $\sim 45^{\circ}$ and the radial intensity profiles passing through the crescent (peaks and troughs inside the ring at $\sim 68$ au; see Appendix \ref{sec:radial_intensity} for more details).

\begin{figure}[h]
    \centering
    \includegraphics[width=0.95\linewidth]{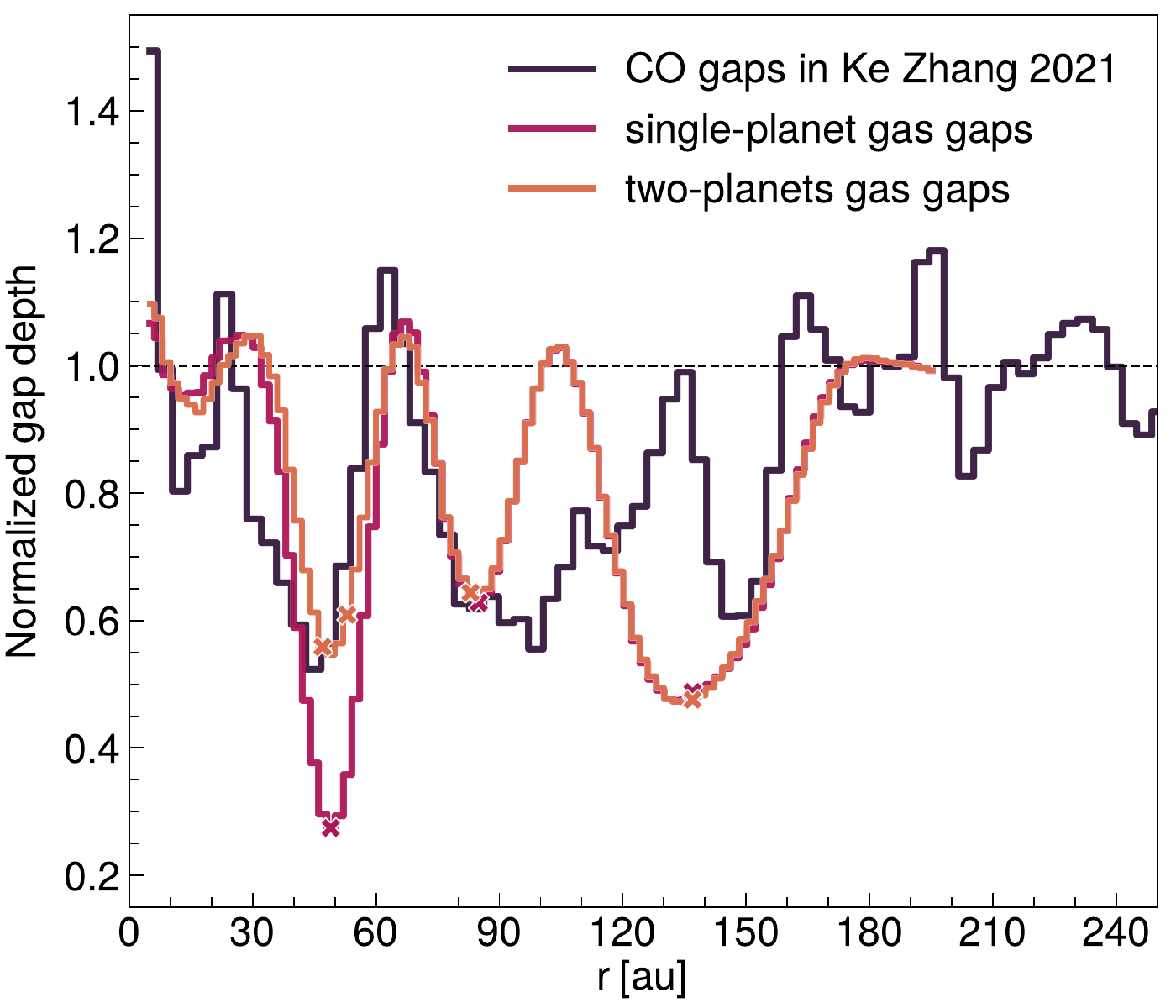}
    \caption{Comparison of column density gap in $\rm C^{18}O\;(2-1)$ line observation found by \citet{Zhang2021} with gas gaps derived from surface density profiles in this work.}
    \label{fig:CO_Comparison}
\end{figure}

\subsection{Dynamical behavior}

Figure \ref{fig:Density_Distribution} compares the gas and dust density distribution for different dust grain sizes. All the dust species show a clear shared gap between the two inner planets. In the largest size, the co-orbital regions of each one display an overdensity at $L_5$, but the more prominent substructure belongs to the second planet. This output has taken into account two conditions. The choice of planetary masses in the inner system must be lower for the body orbiting the substructure, and the presence of the two outer planets. If we neglect either, the Lagrange points can still trap dust, but the mass distribution may change to make some $L_4$ or the inner $L_5$ the most prominent. As shown in \citet{Garrido-Deutelmoser}, this evolution is highly dynamic and irregular so that the overdensities around $L_4$ and $L_5$ constantly change. However, the final morphology in our configuration comes from the early stages of evolution.

Theoretically, the stable Lagrange point $L_5$ is located at $60^{\circ}$ from the planet at its trailing position. However, our model shows that the center of the crescent is slightly shifted and can vary between $65^{\circ}$ and $85^{\circ}$ for reason we still do not understand and deserve further investigation. Accordingly, the position of the proposed planet will be also slightly shifted from the Lagrange point. Finally, we observe that azimuthal extent of the crescent remains roughly constant and equal to $\sim 45^{\circ}$ similar to the observations.

\subsection{Behavior for different dust species}

\begin{figure}[h]
    \centering
    \includegraphics[width=0.95\linewidth]{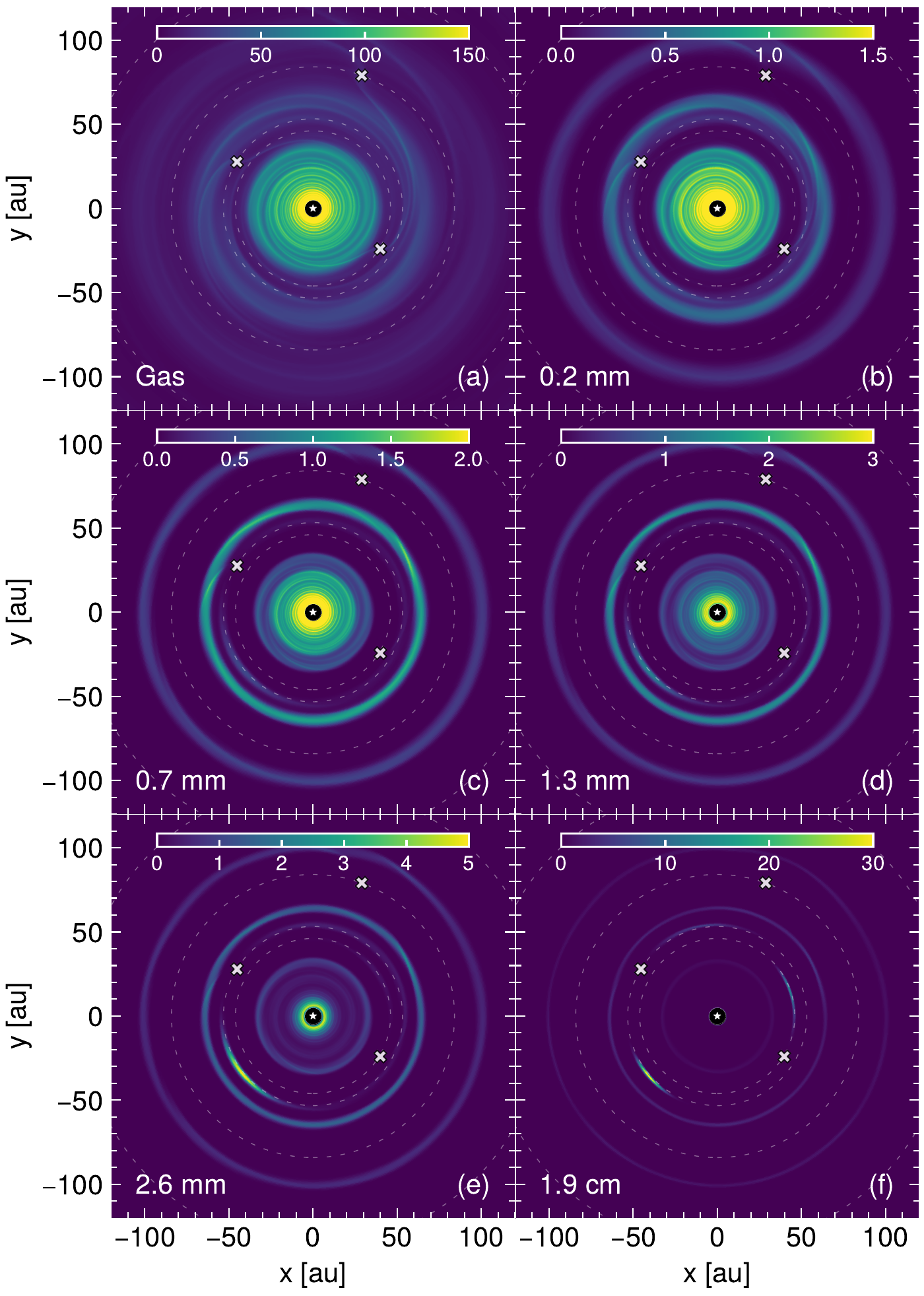}
    \caption{Face-on gas and dust surface density $\Sigma$ from the hydrodynamic model after $\approx 4.8\times 10^5$ yrs (2000 orbits at 48 au). The panels correspond to different fluids. The crosses denote the position of the planets and the white dashed lines indicate their orbits. The disk rotates in a clockwise direction.
    }
    \label{fig:Density_Distribution}
\end{figure}

In our fiducial model, the vortensity for $L_5$ of the outer planet is stronger than that of the inner one (not shown). Therefore, the dust accumulation is generally expected to be greater around the orbit of the outer planet for all sizes. This is especially true for small sizes that are well coupled to the gas and librate with larger amplitudes around $L_5$ (panels b and c in Fig. \ref{fig:Density_Distribution}). As the size of the grains increases (Stokes numbers approach unity, panels d to f), the dust distribution becomes compact toward the center of the Lagrange point \citep{Montesinos2020} and we can even see some tenuous accumulation around the inner planet's $L_5$ point at 1.9 cm (panel f).

\begin{figure*}[t]
    \centering
    \includegraphics[width=0.95\linewidth]{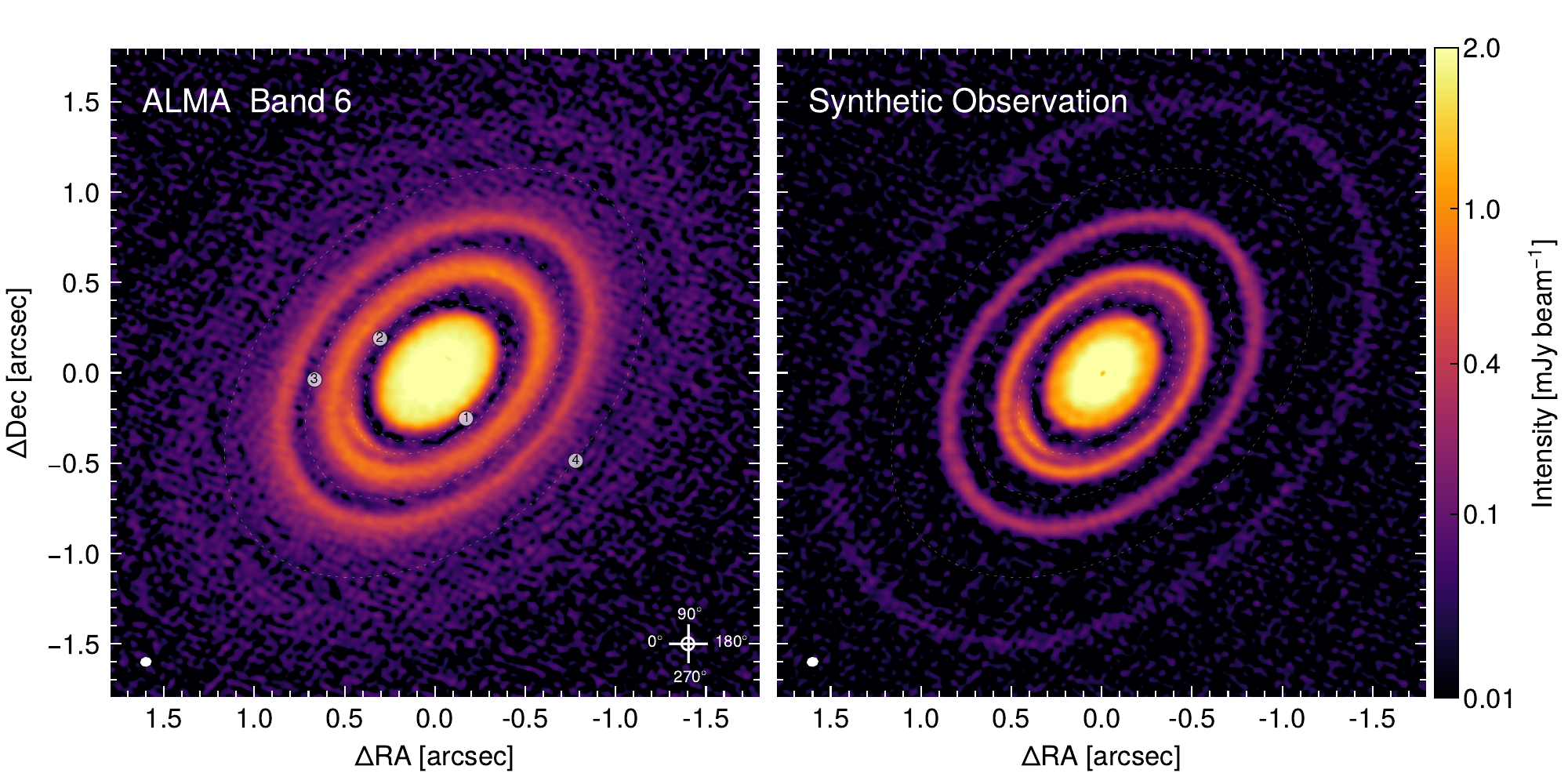}
    \caption{Band 6 ($\lambda \sim$1.25 mm) comparison between dust continuum image from ALMA observation \citep{Isella2018} and our synthetic observation after $\approx 4.8\times10^{5}$ yrs. The synthesized beam are the same for both images (0.038$''$ $\times$ 0.048$''$, 82.5$^{\circ}$), represented by white ellipse at the bottom left corner for each image. The synthetic image is projected with an inclination $i = 46^{\circ}$ and a position angle PA$=133^{\circ}$. The enumerated white dots indicate the position of potential planets associated with: (1) and (4) resonance angles in Laplace chains (see \S\ref{subsec:predictions}), (2) Lagrange point $L_5$ from the simulation, (3) velocity kink reported by \citet{Pinte2020}. The white dashed lines denote the orbits of planets in the simulation. The right bottom Cartesian coordinate describe the prescription to estimate the azimuthal angles. The disk rotates in a clockwise direction.}
    \label{fig:ALMA_Comparison}
\end{figure*}

\section{A Laplace resonance chain}

Our fiducial simulation has 4 planets with period ratios $P_{2}/P_{1}=(54/46)^{3/2}=1.27$, $(84.5/54)^{3/2}=1.96$, and $(137/84.5)^{3/2}=2.06$. Therefore, the three outer planets lie near a $1:2:4$ commensurability, which becomes nearly exact (within $1\%$) if planet 3 changes from 84.5 au to 86 au. This fact begs the question of whether disk-driven migration may have placed the planets in their current, near resonant, orbits\footnote{We note that a disk-driven migration may also lead to offsets from the exact commensurabilities by either wake-planet interactions \citep{Baruteau2013}, disk-driven precession \citep[e.g.,][]{Tamayo2015} or resonant repulsion \citep[e.g.,][]{Papaloizou2011}. }. 

From Figure \ref{fig:Depth_Comparison}, the planets carved relatively shallow gaps, so we may estimate the rate of orbital migration following \citealt{Kanagawa2020} as:
\begin{eqnarray}
\tau_{a} \equiv \left| \frac{a}{\dot{a}} \right | &\simeq& \left(\frac{M_\star}{M_{\rm p}}\right) \left(\frac{M_\star}{\Sigma_{\rm min} a_{\rm p}^2}\right) \frac{h_{\rm p}^2}{\Omega_{\rm K,p}},\nonumber\\ 
&\simeq& 0.4 \; \mbox{Myr} \left(\frac{10 \;{\rm gr\;cm^{-2}}}{\Sigma_{\rm min}}\right) \left(\frac{100 \;{\rm au}}{a_{\rm p}}\right)^{1/2} \nonumber\\
&& \left(\frac{2M_{\odot}}{M_{\star}}\right) \left(\frac{1M_{\rm J}}{M_{\rm p}}\right) \left(\frac{h_{\rm p}}{0.1}\right)^2.
\label{eq:tau_a}
\end{eqnarray}
where $\Sigma_{\rm min}$ corresponds to the local density at the base of the density gap. Using the fiducial planetary parameters and $M_\star = 1.9 \, M_\odot$, a fix aspect ratio $h=0.1$ and the surface density constraints from \citet[][Table 5 therein]{Zhang2021}, we observe that the migration timescales are all comparable to the age of the system making migration a plausible scenario.

As a proof of concept, in Figure \ref{fig:Laplace} we show an N-body integration using \texttt{REBOUND} \citep{Rebound2012} and prescribing the damping timescales $\tau_a$ and $\tau_e\equiv e/|\dot{e}|$ for each planet in order to mimic planet-disk interactions in the \texttt{REBOUNDx} library \citep{Tamayo2019}. We set $\tau_a/\tau_e=100$ with $\tau_a$ computed using Eq.~\ref{eq:tau_a}, see values for the orbital decay timescales in Table \ref{tab:models_hd163296}. We begin the simulations with the planets further away from their current positions and let them migrate due to their interaction with the gaseous component of the disk, 
where kinematic evidence has been detected in the gas at $\sim 260$ au \citep{Pinte2020,Teague2021}, and extensions in CO (2-1) up to $\sim 500$ au \citep{Zhang2021}. 
The evolution shows that all planet pairs are captured into a long resonant chain after $\sim 0.5 \times 10^6$ yrs. These correspond to a two-body 4:3 mean-motion resonance (MMR) for the innermost planets, and a double 2:1 - 2:1 MMR for the two outer pairs, finally leading to the libration of the following three-body angles:
\begin{eqnarray}
\phi_{123}&=&3\lambda_1-5\lambda_2+2\lambda_3, \mbox{ and}\\
\phi_{234}&=&2\lambda_2-6\lambda_3+4\lambda_4.
\end{eqnarray}

Both $\phi_{123}$ and $\phi_{234}$ have small-amplitude libration ($\sim 2^\circ$) and their libration centers are $187^\circ$ and $218^\circ$, respectively. Note that when the four planets reach the reported semi-major axis at approximately the same time $\sim 10^6$ yrs (gray vertical line in panel b), they are already captured in the two- and three-planet resonances, showing that the proposed configuration with our hydrodynamical simulations presented in the previous sections is possible.

\subsection{Predicting the position angle (PA) of the planet candidates using 3-body resonances}\label{subsec:predictions}

Because the orbits of planets $i$ are coplanar and nearly circular ($e_i \lesssim 0.07$), the mean longitudes $\lambda_i$ are close to the true longitudes $\varpi_i+f_i$ and will likely librate around the same angles. Defining an arbitrary reference frame, rotated by ${\rm PA}_0$ we write the position angles (PAs) as ${\rm PA}_i=\varpi_i+f_i+{\rm PA}_0$, and define the following three-body angle, frame-independent\footnote{Since the critical angles satisfy the D'Alembert property, these combinations are independent of the reference frame.}, combinations:
\begin{eqnarray}
{\rm PA}_{123}&=&3{\rm PA}_1  -5{\rm PA}_2 +2{\rm PA}_3, ~~\mbox{ and}\\ {\rm PA}_{234}&=&2{\rm PA}_2 -6{\rm PA}_3 +4{\rm PA}_4.
\end{eqnarray}

\begin{figure}[h]
    \centering
    \includegraphics[width=0.95\linewidth]{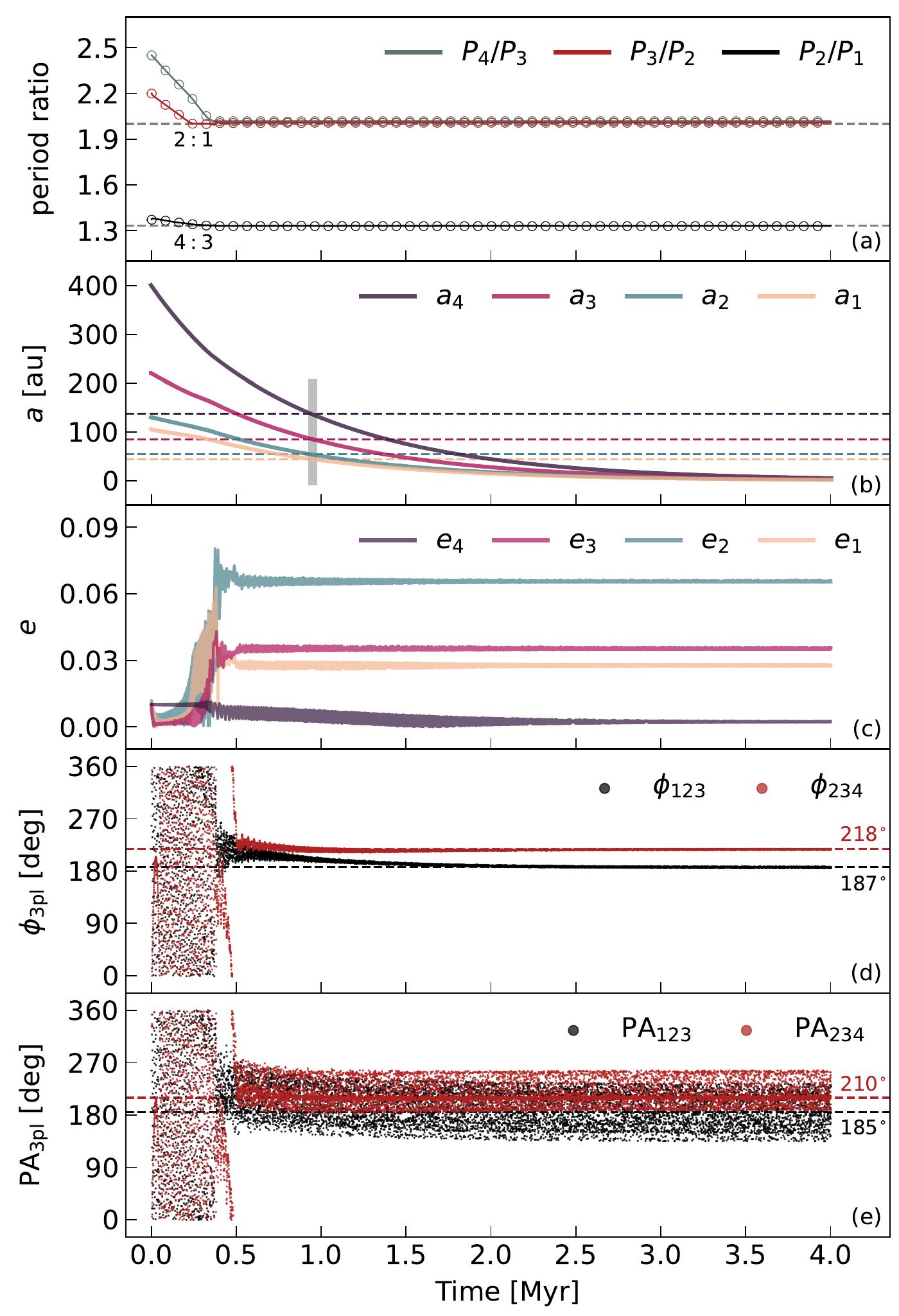}
    \caption{Potential migratory history of the four-planet system locking the planets in a long orbital resonance chain leaving the outer three planets near a consecutive 2:1 commensurability and the innermost pair near 4:3 (panels a and b). The eccentricities remain small after the capture (panel c) and the three-body resonant angles
    $\phi_{123}=3\lambda_1-5\lambda_2+2\lambda_3$ and  $\phi_{234}=2\lambda_2-6\lambda_3+4\lambda_4$ undergo small-amplitude libration (panel d). The bottom panel exhibits the corresponding combinations of the position angles in the system: ${\rm PA}_{123}=3{\rm PA}_1  -5{\rm PA}_2 +2{\rm PA}_3$ and ${\rm PA}_{234}=2{\rm PA}_2 -6{\rm PA}_3 +4{\rm PA}_4$. }
    \label{fig:Laplace}
\end{figure}

From panel (e) in Figure \ref{fig:Laplace}, we observe that these angles librate around ${\rm PA}_{123}\sim 185^\circ$ and ${\rm PA}_{234}\sim 210^\circ$ similar to $ \phi_{123}$ and $ \phi_{234}$, but with larger amplitudes (near $100^\circ$ in both cases). This is expected due to the non-zero eccentricities.

Assuming that we know two angles, say ${\rm PA}_2$ due to the crescent and ${\rm PA}_3$ due to a velocity kink, we can use the above relations to constrain ${\rm PA}_1$ and ${\rm PA}_4$ as:
\begin{eqnarray}
{\rm PA}_1 &\sim & {\textstyle \frac{1}{3} {\rm PA_{123}}} + {\textstyle \frac{5}{3} {\rm PA_{2}}} - {\textstyle \frac{2}{3} {\rm PA_{3}}}, ~~\mbox{ and}\\
{\rm PA}_4 &\sim & {\textstyle \frac{1}{4} {\rm PA_{234}}} - {\textstyle \frac{1}{2} {\rm PA_{2}}} + {\textstyle \frac{3}{2} {\rm PA_{3}}}.
\end{eqnarray}

Our model indicates that planet 2 (54 au) is $\sim 75^\circ\pm10^\circ$ ahead of the crescent center\footnote{Due to the clockwise rotation of the disk, our angle convention PA is given the coordinate axes at the bottom of Figure \ref{fig:ALMA_Comparison}.}, which corresponds to $\rm PA_2 \simeq 32^\circ$. In addition, as reported by \citet{Pinte2020}, the planet 3 at 86 au has $\rm PA_3 \simeq 357^\circ$ associated with a velocity kink. Thus, our model predicts that the planets at $46$ au and 137 au should have position angles of $\rm PA_1 \sim 237^{\circ}$ and $\rm PA_4 \sim 212^{\circ}$ respectively (see the left panel in Figure \ref{fig:ALMA_Comparison}).

Quite recently\footnote{After the submission of our manuscript to the journal.}, \citet{Alarcon2022} localized strong kinematic deviation in C I line emission. The position of this structure lies inside the gap at 48 au, which azimuthally coincides with our predicted planet 1 at $\rm PA_1 \sim 237^{\circ}$. We note that this predicted planet differs from the one proposed by \citet{Isella2018} and incorrectly quoted by \citet{Alarcon2022} as coincident with the outflow. The reason is that the disk rotates in a clockwise direction so the proposed planet invoked to explain the crescent as a $L_5$ feature, similar to \citet{Rodenkirch2021}, will actually show ahead of the crescent. In this way, the C I deviation cannot be explained by a co-rotational planet that is also responsible for the crescent, unless the dust accumulation around $L_4$ becomes more prominent than that of $L_5$, which is unlikely \citep{Rodenkirch2021,Garrido-Deutelmoser}.

\begin{deluxetable}{cccc}
\label{tab:models_hd163296}
\tablecaption{Migration rate estimates}
\startdata
--- & --- & --- & ---\\
$a_{\textrm{p}}$ & $M_{\textrm{p}}$ & $\Sigma_{\textrm{min}} \textrm{ [gr/cm/cm]}$ & $\tau_{\textrm{a}}$ \textrm{ [Myr]} \\
46    au  & 85  $M_\oplus$ & 12        & 1.8   \\
54    au  & 60  $M_\oplus$ & 19        & 1.5   \\
84.5  au  & 127 $M_\oplus$ & 9.3       & 1.1   \\
137   au  & 317 $M_\oplus$ & 4.2       & 0.8   \\
\enddata
\tablecomments{The values of $\Sigma_{\rm min}$ are taken from Table 5 in \citet{Zhang2021}, except for the innermost one provided by our model.}
\end{deluxetable}

\subsection{Resonances in other systems}

We remark that resonances may be a common outcome in these young systems, including the embedded planets in PDS 70 \citep{Bae2019}, as well as young, but disk-free systems, like HR 8799 also in a long resonance chain involving four planets \citep{Gozdziewski2020}.

Similar to our work, a compact multi-planet system has been proposed using the axisymmetric dust gaps and rings of HL Tau \citep{ALMA2015}, where a resonant configuration may promote the system's dynamical stability \citep{Tamayo2015}. In our case, we use not only the system's migration history and dust rings and gaps, but also add the constraints from the crescent shape structure and the CO gas emission.

\section{Conclusions}

We have provided a global model for HD 163296 with four planets (semi-major axes in the range of $40-140$ au) that can reproduce the rings and gaps in the dust continuum and the shallow gaps in the gas constrained by the CO emission. A key ingredient in our model is the presence of two sub-Saturn-mass planets near the 4:3 resonance opening  the gap at $\sim 48$ au, where the crescent corresponds to the $L_5$ Lagrange point of the outer planet at 54 au. 

We show that the four-planet system may be part of a long resonance chain with the inner two in a 4:3 MMR and the outer three in a 1:2:4 Laplace resonance chain, consistent with a history of convergent migration within the disk. Our proposed three-body resonances allow to relate  the planetary radial and angular positions, and based on the crescent location at 55 au and the proposed location by \citet{Pinte2020} for the planet at $\simeq 86$ au, our model predicts two planets: i) a sub-Saturn at 46 au and $\rm PA \sim 237^{\circ}$; ii) a Jovian at 137 au  and $\rm PA \sim 212^{\circ}$(Figure \ref{fig:ALMA_Comparison}). 

Overall, our work shows that tightly-spaced planetary systems, often found at small orbital distances in transiting surveys, may leave detectable imprints in protoplanetary disks at much larger separations.

\begin{acknowledgements}
\emph{Acknowledgements} The authors would like to thank Andrew Youdin, Kaitlin Kratter, Diego Mu\~noz, Matt Russo, Pablo Benítez-Llambay, Sim\'on Cassasus, and Ximena S. Ramos for helpful discussions that improved the quality of this work and Juan Veliz for his support with the cluster logistics. Finally we thank the anonymous reviewer for the thorough and useful report.
J.G. acknowledge support by ANID, -- Millennium Science Initiative Program -- NCN19\_171 and FONDECYT Regular grant 1210425. The Geryon cluster at the Centro de Astro-Ingenieria UC was extensively used for the calculations performed in this paper. BASAL CATA PFB-06, the Anillo ACT-86, FONDEQUIP AIC-57, and QUIMAL 130008 provided funding for several improvements to the Geryon cluster.
C.P. acknowledges support from ANID Millennium Science Initiative-ICN12\_009, CATA-Basal AFB-170002, ANID BASAL project FB210003, FONDECYT Regular grant 1210425, CASSACA grant CCJRF2105, and ANID+REC Convocatoria Nacional subvencion a la instalacion en la Academia convocatoria 2020 PAI77200076.
C.C. acknowledges FNRS Grant No. F.4523.20 (DYNAMITE MIS-project). 
V.V.G. acknowledges support from FONDECYT Regular 1221352, ANID project Basal AFB-170002, and ANID, – Millennium Science Initiative Program – NCN19\_171.
K.Z. acknowledges the support of the Office of the Vice Chancellor for Research and Graduate Education at the University of Wisconsin – Madison with funding from the Wisconsin Alumni Research Foundation.
\end{acknowledgements}

\software{\textsc{Fargo3D} \citep{Benitez2019}, \textsc{Numpy} \citep{vanderWalt2011}, \textsc{Matplotlib} \citep{Hunter2007}. \textsc{Rebound} \citep{Rebound2012}, \textsc{ReboundX} \citep{Tamayo2019}, RADMC3D \citep{Dullemond2012}.}


\bibliography{sample631}{}
\bibliographystyle{aasjournal}



\appendix

\section{GAS GAP CALCULATION} \label{sec:appendix_fit}

\begin{figure}[h]
    \centering
    \includegraphics[width=0.95\linewidth]{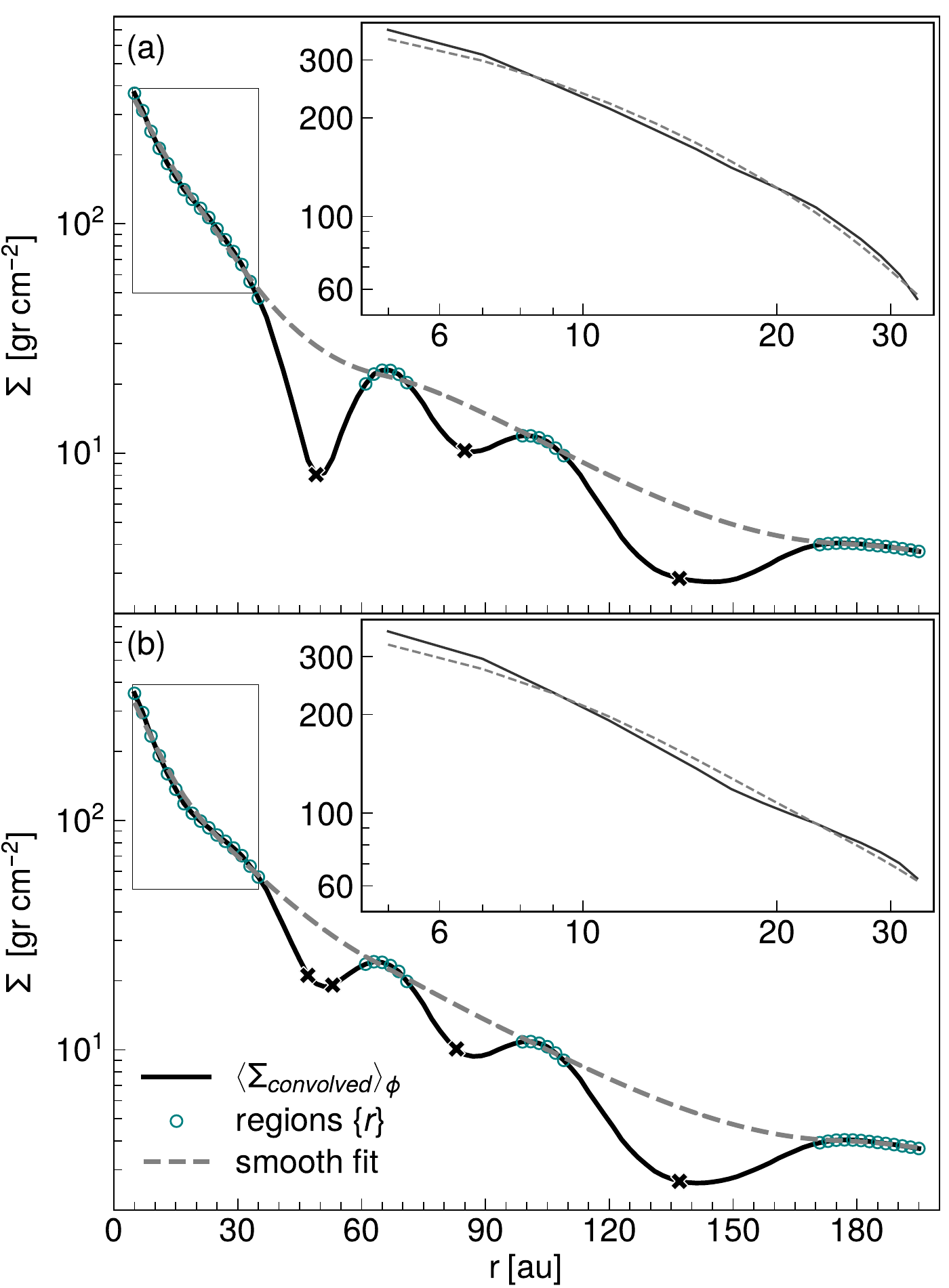}
    \caption{Black line denote the convolved surface density profile of models and grey dashed line the respective smooth function. Panel (a) and (b) represent the single-planet case and two-planet case respectively. The crosses indicate the position of the planets.}
    \label{fig:smooth_fit}
\end{figure}

In \citet{Zhang2021} a smooth function was subtracted from $\rm N_{CO}$ column density profiles to better characterize substructures in the residual values. To compare these results with our models, we follow the same procedure. First, the $\Sigma$ maps from hydrodynamic simulations were convolved with a circular Gaussian beam of $0.15''$, which has the same size as MAPS CO (2-1) line observations. Then, their azimuthally averaged profiles were interpolated every 2 au. In addition, the radial region $\{r \;{\rm[au]}: 0<r_0<35 , 59<r_1<72 , 98<r_2<110 , r_3>170\}$ was selected to describe the gaps. Both were taken as input for the \textsc{smoothfit}\footnote{https://pypi.org/project/smoothfit/} module.
The Figure \ref{fig:smooth_fit} shows the outputs of smoothed profile represented by the grey dashed line and the convolved surface density profile in black lines. The cyan dots denote the regions in which the function acts. The Figure \ref{fig:CO_Comparison} shows the residual between lines to provide a reasonable comparison with CO gaps observations.

\section{Radial intensity}\label{sec:radial_intensity}

\begin{figure}
        \centering
        \includegraphics[width=0.95\linewidth]{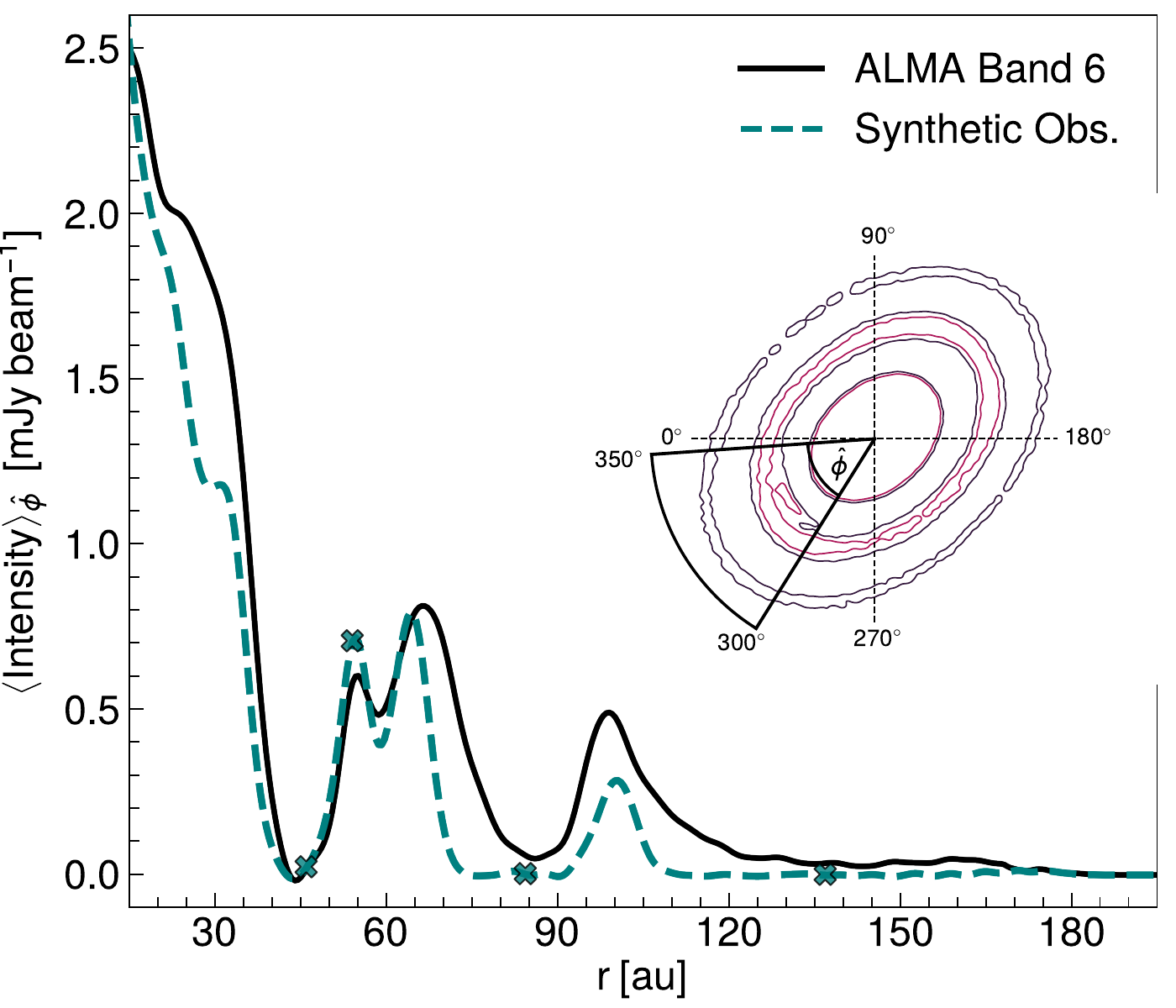}
        \caption{Azimuthally averaged intensity profiles for synthetic and ALMA observations after $\approx 4.8 \times 10^5$ yrs. The crosses denote the semi-major axes of the planets. The insert shows the ALMA observation in Band 6 with contours that reproduce the crescent and rings as well as the angular slice used for the azimuthal average denoted by $\hat{\phi}$. }
        \label{fig:radia_intensity}
    \end{figure}

We quantify the intensity around the substructure region of our synthetic model with the ALMA observation. First, we deproject the images obtaining a face-on view to convert them to polar coordinates and then generate a radial profile by taking the azimuthal average between PA of $300^\circ$ and $350^\circ$. This extension fully covers the emission from the crescent. The results are shown in the Figure \ref{fig:radia_intensity}, which is accompanied by a diagram showing the angular slice.

Figure \ref{fig:radia_intensity} show that radial intensity through the crescent region reaches amplitudes higher than those observed by a factor of 1.2 at 55 au. The emission from the substructure is clearly off-centered on the gap and resolved in spatial resolution, showing a gap in intensity between it and the ring. The first ring reproduces the intensities in a good way, while the second is noticeably 1.7 times fainter.

\section{Effect from disk gravity acting on planets}\label{sec:planetdisk}

We briefly test whether turning on the full disk-planet interaction may lead to morphological changes in the structure of the crescent. We recall that in our fiducial simulation (see \S\ref{sec:hydro_sims}), while the disk do feel the planets' gravity, the planets do not feel the disk.  

\begin{figure}
        \centering
        \includegraphics[width=0.95\linewidth]{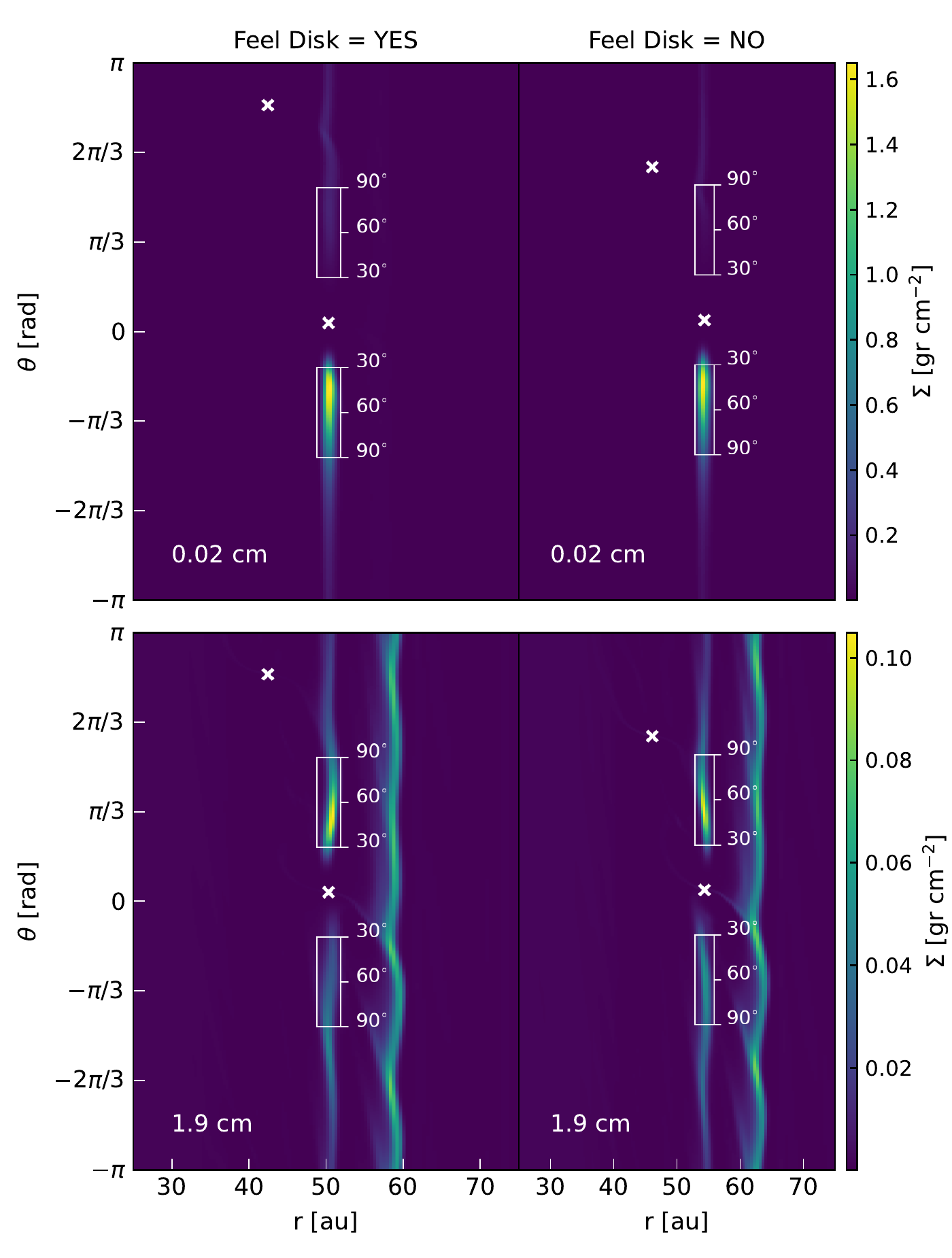}
        \caption{Dust surface density $\Sigma$ for 0.02 cm and 1.9 cm grain sizes. The crosses denote the position of the planets. The upper and bottom white rectangles, indicate the Lagrange points $L_4$ and $L_5$ respect to the outer planet.}
        \label{fig:planetdisk}
    \end{figure}

We perform two-planet simulations considering only the inner planet pair near the 4:3 commensurability (46 au and 55 au) for up to 2000 orbits of the inner planet. The initial density has been reduced by a factor of 100 to avoid significant migration. In figure \ref{fig:planetdisk} we show the density distribution for two dust fluids of 0.02 cm and 1.9 cm grain sizes in two cases: the full disk-planet interaction is considered (left panels, displaying a slight inward migration at the $\sim 10\%$ level), and the disk gravity acting on the planets is ignored (right panels, with no migration). Despite of the slight orbital migration, we do not observe any significant changes regarding the amount and distribution of captured material  at the $L_4$ and $L_5$ Lagrange points.

\end{document}